\documentclass[prl,twocolumn,superscriptaddress,floatfix,noshowpacs,10pt,longbibliography]{revtex4-2}
\usepackage{graphicx,bm,times}
\usepackage{amsmath}
\usepackage{amsfonts}
\usepackage{amssymb}
\usepackage[version=4]{mhchem} 

\usepackage{bm}
\usepackage{color}
\usepackage{times}
\usepackage{xcolor}
\definecolor{gray}{RGB}{110,110,110}
\usepackage[%
  colorlinks=true,
  urlcolor=blue,
  linkcolor=blue,
  citecolor=blue
  ]{hyperref}

\begin{document}

\title{Anisotropic magnetic interactions in a candidate Kitaev spin liquid \\ close to a metal-insulator transition}

\author{Zeyu~Ma}
\affiliation{Clarendon Laboratory, Department of Physics,
	University of Oxford, Parks Road, Oxford OX1 3PU, UK}

\author{Danrui~Ni}
\affiliation{Department of Chemistry, Princeton University,
Princeton, NJ 08544, USA}

\author{David~A.~S.~Kaib}
\affiliation{Institut für Theoretische Physik, Goethe-Universität Frankfurt, 60438, Frankfurt am Main, Germany}

\author{Kylie~MacFarquharson}
\affiliation{Clarendon Laboratory, Department of Physics,
	University of Oxford, Parks Road, Oxford OX1 3PU, UK}

\author{John~S.~Pearce}
\affiliation{Clarendon Laboratory, Department of Physics,
	University of Oxford, Parks Road, Oxford OX1 3PU, UK}

\author{Robert~J.~Cava}
\affiliation{Department of Chemistry, Princeton University,
Princeton, NJ 08544, USA}

\author{Roser~Valent\'\i}
\affiliation{Institut für Theoretische Physik, Goethe-Universität Frankfurt, 60438, Frankfurt am Main, Germany}

\author{Radu~Coldea}
\affiliation{Clarendon Laboratory, Department of Physics,
University of Oxford, Parks Road, Oxford OX1 3PU, UK}

\author{Amalia~I.~Coldea}
\email[corresponding author:]{amalia.coldea@physics.ox.ac.uk}
\affiliation{Clarendon Laboratory, Department of Physics,
University of Oxford, Parks Road, Oxford OX1 3PU, UK}

\date{\today}

\begin{abstract}
In the Kitaev honeycomb model, spins coupled by strongly-frustrated anisotropic interactions do not order at low temperature but instead form a quantum spin liquid with spin fractionalization into Majorana fermions and static fluxes. The realization of such a model in crystalline materials could lead to major breakthroughs in understanding entangled quantum states, however achieving this in practice
is a very challenging task. The recently synthesized honeycomb material RuI$_3$ shows no long-range magnetic order down to the lowest probed temperatures and has been theoretically proposed as a quantum spin liquid candidate material on the verge of an insulator to metal transition. Here we report a comprehensive study of the magnetic anisotropy in un-twinned single crystals via torque magnetometry and detect clear signatures of strongly anisotropic and frustrated magnetic interactions.
We attribute the development of sawtooth and six-fold torque
signal to strongly anisotropic,
bond-dependent magnetic interactions  by comparing to
theoretical calculations.
As a function of magnetic field strength at low temperatures,
torque shows an unusual non-parabolic dependence suggestive of a proximity to a field-induced transition.
Thus, RuI$_3$, without signatures of long-range magnetic order, displays key hallmarks of an exciting new candidate for extended Kitaev magnetism
with enhanced quantum fluctuations.
\end{abstract}

\maketitle

\begin{figure}[htbp]
\centering
\includegraphics[trim={0cm 0cm 0cm 0cm},width=0.95\linewidth]{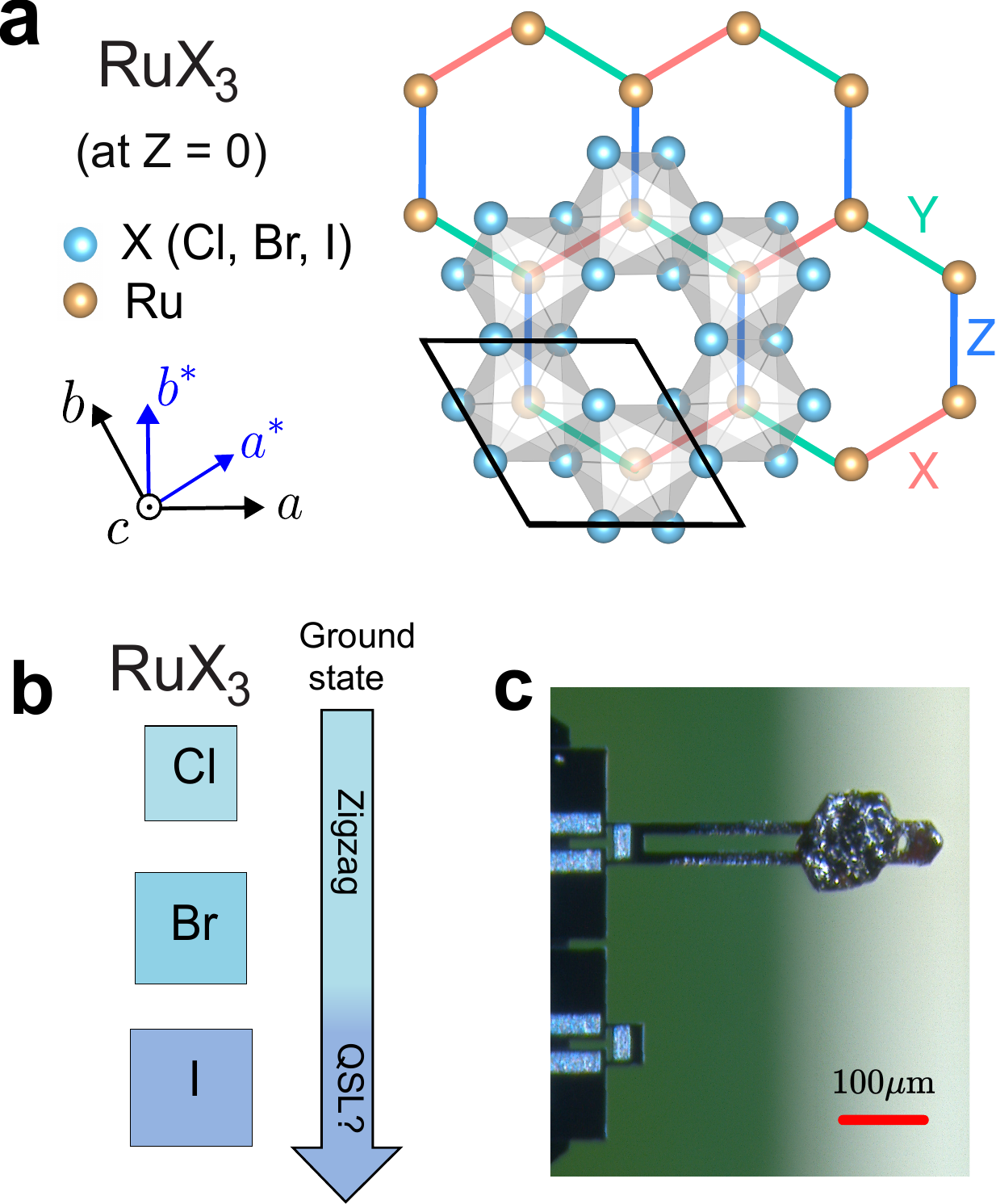}
\caption{{\bf Crystal structure and phase diagram of RuX$_3$ halides.} (a) Projection of a honeycomb layer of edge-sharing RuX$_6$ octahedra on the $ab$ plane (at $z=0$) with axes orientation in the hexagonal setting for the $R\overline{3}$ (No.~148) space group~\cite{Ni2022}. Solid black rhombus outlines the hexagonal unit cell with the $b^*$ axis along a Ru-Ru bond. The different colours of the Ru-Ru bonds  indicate the Ising character of the interactions in a pure Kitaev model. (b) Summary of the magnetic ground state of RuX$_3$ halides with X=Cl, Br and I with increasing unit cell volume indicating the low temperature zigzag antiferromagnetic order for X=Cl and Br, and proposed quantum spin liquid (QSL) for X=I, respectively. (c) A single crystal of RuI$_3$ mounted on a piezocantilever. }
\label{Fig1}
\end{figure}

The Kitaev model with bond-dependent Ising interactions on a two-dimensional honeycomb lattice (see Fig.~\ref{Fig1}(a)) has an exactly solvable quantum spin liquid ground state \cite{kitaev2006anyons,Takagi2019,Trebst2022}. This can be intuitively thought of as a quantum superposition of many classical configurations, having a third of the bonds minimizing their energy by aligning the spins along their respective Ising bond axis, with the rest of bonds gaining zero energy.
In real materials, achieving a pristine Kitaev ground state is challenging due to the presence of both isotropic and anisotropic interactions that extend beyond the bond-dependent Ising exchange. These additional interactions can give rise to a variety of long-range ordered states or, intriguingly, different types of quantum spin liquids
\cite{Rau2014,winter2017models,luo2021gapless,gohlke2022extended}.

Among proposed potential Kitaev platforms \cite{Jackeli2009}, $\alpha$-RuCl$_3$ with Ru$^{3+}$ ions in a $d^5$ electronic configuration and a rock-salt-related honeycomb structure, has been intensively studied. Despite the fact that orders magnetically into a zigzag antiferromagnetic structure~\cite{johnson2015} below 7~K, it displays several features in the spin dynamics consistent with dominant Kitaev interactions \cite{banerjee2017neutron,winter2017breakdown}. This compound is broadly described by the $K\!-\!J\!-\!\Gamma\!-\!\Gamma'$ model for the effective spin-1/2 Ru$^{3+}$ moments, where the Kitaev interaction originates from chlorine-mediated exchange through edge-shared octahedra arranged on a honeycomb lattice, see Fig.~\ref{Fig1}(a). In the presence of an external magnetic field, the magnetic order is suppressed above 7–8~T and unusual magnetic excitations and thermal transport have been reported. Some studies suggested a potential half-integer quantized thermal Hall effect due to the possible existence of chiral Majorana edge modes expected in the pure Kitaev model \cite{kasahara2018majorana,yokoi2021half}, whereas other studies emphasized the effects induced by structural layer stacking faults \cite{Bruin2022}, topological magnons \cite{Czajka2023} or phonons \cite{lefranccois2022}.

Chemical pressure via isoelectronic substitution  allows the exploration of the magnetic phase diagram via tuning the magnetic interactions (Fig.~\ref{Fig1}(b)). Isostructural RuBr$_3$ shows onset of zigzag magnetic order at an even higher temperature of 34~K than $\alpha$-RuCl$_3$ \cite{Imai2022}, whereas no magnetic transition was detected in RuI$_3$ so far \cite{Nawa2021,Ni2022}.
Here we report torque measurements on high-quality un-twinned single crystals of RuI$_3$, which observe a strong magnetic torque response attributed to the magnetism of localized Ru$^{3+}$ magnetic moments. Furthermore, the angular dependence of torque for magnetic field in three orthogonal crystallographic planes provide direct evidence that the magnetic interactions are strongly anisotropic and highly-frustrated, making RuI$_3$ a strong candidate to realize unconventional cooperative magnetism in the absence of long-range order.

\begin{figure*}[htbp]
\centering
\includegraphics[trim={0cm 0cm 0cm 0cm},width=0.90\linewidth]{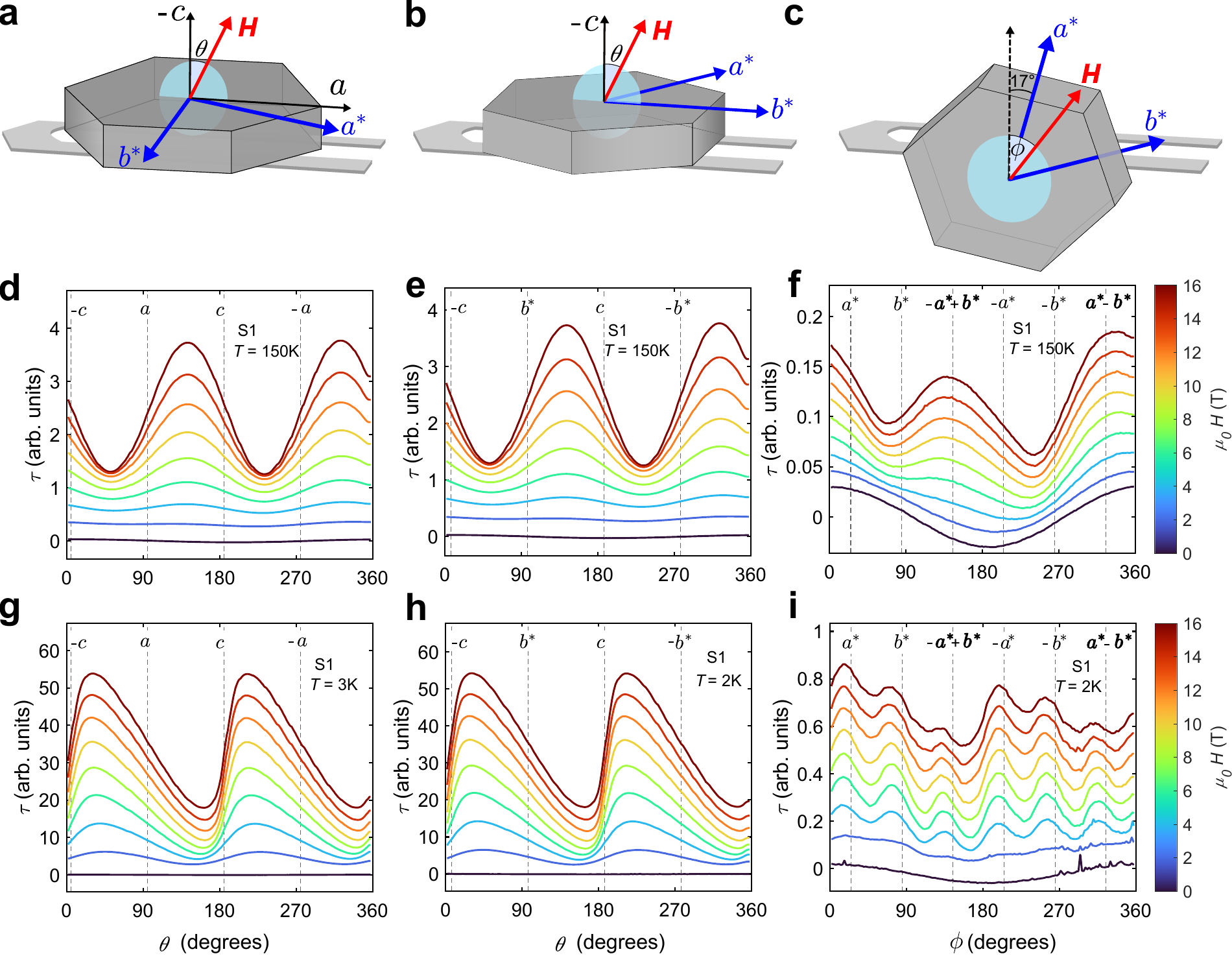}
\caption{
{\bf Angular dependence of torque for magnetic field orientation in three orthogonal crystallographic planes, at different fixed temperature and various magnetic field strengths.} (a-c) Schematic diagram of the hexagonally-shaped samples mounted on the piezocantilever in three orientations. Axes labels are with reference to the crystal structure illustrated in Fig.~\ref{Fig1}(a). The light blue disk shows the plane in which the magnetic field orientation changes relative to the crystal axes, where in (a,b) $\theta=0^\circ$ is close to $H\parallel -c$ and in (c) $\phi=17^\circ$ is close to $H\parallel a^*$. (d-i) Angular dependence of torque for fixed field strength increasing from bottom (0~T) to top (16~T), with curves offset vertically for clarity and colour-coded in the right-hand colorbars. Top axes labels and vertical dashed lines indicate when the field direction is close to reference crystal axes in the rotation plane. Middle row shows data at high temperature (150~K) and the bottom row at base temperature (2~ or 3~K).
}
\label{Fig2}
\end{figure*}

\begin{figure*}[htbp]
\centering
\includegraphics[trim={0cm 0cm 0cm 0cm},width=0.93\linewidth] {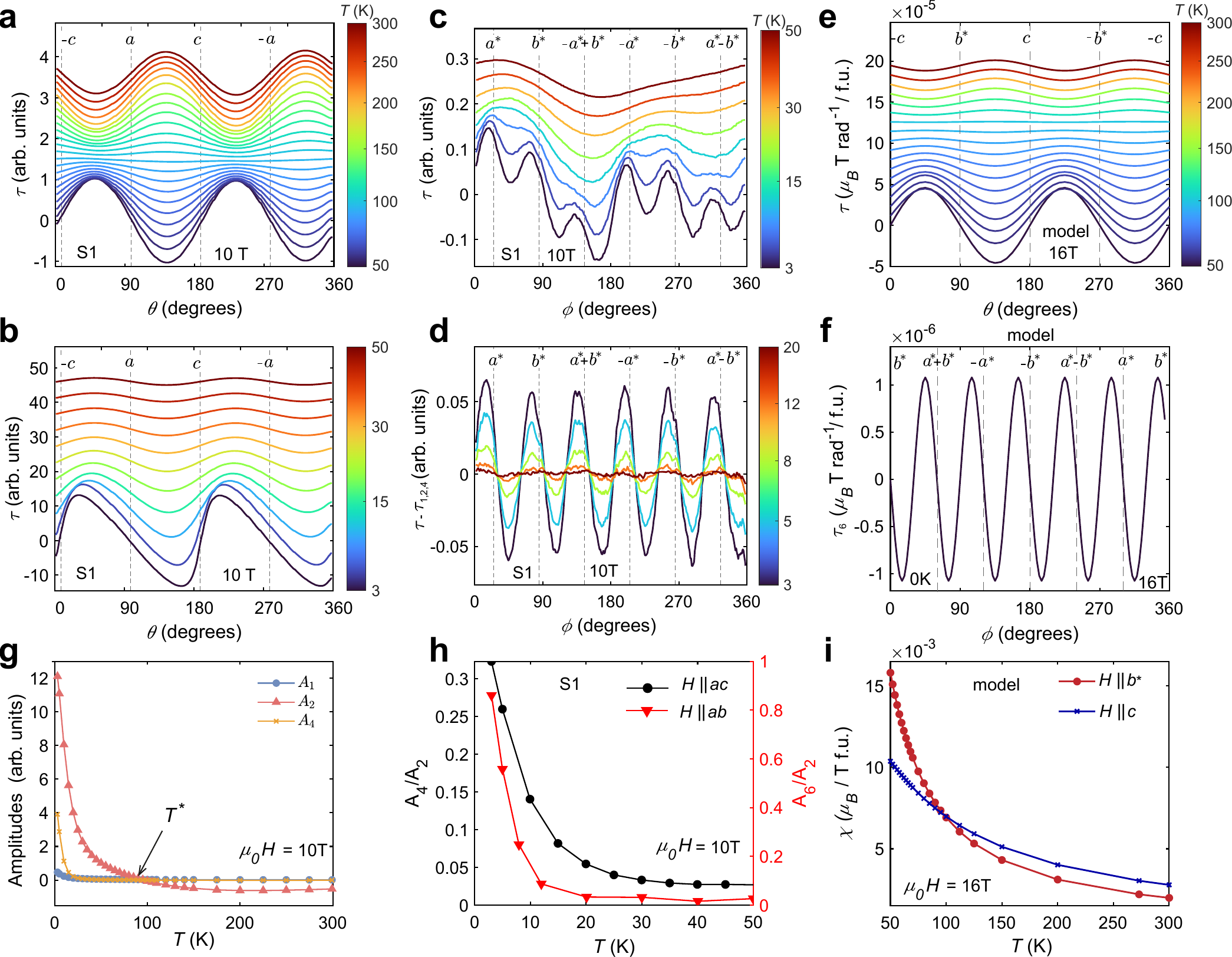}
\caption{
{\bf Angular dependence of torque as a function of temperature.} (a,b) Angular dependence for magnetic field of 10~T in the $ac$ plane. Curves correspond to fixed temperatures and are offset vertically in order of increasing temperature. (c) Same as (b) but for a magnetic field in the $ab$ plane. (d) Angular dependence of the torque after subtracting different Fourier components due to sample weight ($\tau_1$) and small misalignment ($\tau_1$ and $\tau_4$),
as described in the text and shown in Fig.~S4(c) in the SM \cite{SM}.
In (a-d) the top axes labels and vertical dashed lines indicate when the applied field direction is close to special crystallographic axes. (e) and (f) Calculated angular dependence of torque for field normal to or in the honeycomb layers, to be compared to experimental data in panels (a) and (d), respectively. (g) Extracted temperature-dependence of the Fourier components $A_{1}$ (filled circles), $A_{2}$ (filled triangles) with a sign change at $T^*$ (arrow), and $A_{4}$ (crosses). (h) Temperature dependence of the Fourier amplitudes ratio $A_4/A_2$ in the $ac$ plane (solid circles, left axis) characterising the degree of distortion of the angular dependence of torque from a pure sinusoid. Red triangles (right axis) show the temperature-dependence of the $A_6/A_2$ ratio of the Fourier amplitudes in the $ab$ plane, which characterizes the magnitude of the in-plane six-fold modulation relative to the non-intrinsic out-of-plane contribution due to a small $c$-axis misalignment. (i) Theoretical predictions for the temperature dependence of the different components of the magnetic susceptibility at 16~T.
}
\label{Fig3}
\end{figure*}

{\bf Magnetic anisotropy in the plane normal to the honeycomb layers.}
We first characterise the torque for magnetic field in the $ac$ plane normal to the honeycomb layers
(Fig.~\ref{Fig2}(a)), as a function of increasing magnetic field strength. At high temperatures (150~K) the angular dependence of torque (Fig.~\ref{Fig2}(d)) is well-described by a functional form $\tau(\theta) = A_2\sin 2(\theta-\theta_2)$, where $\theta_2$ is a small angular offset from the $H\! \parallel\! c$ orientation. The negative torque gradient for $H\!\parallel\! c$ indicates that this orientation is the {\em stable} equilibrium in this plane, {\it i.e.} the susceptibilities order is $\chi_c>\chi_a$. Very similar behaviour is observed in torque measurements for magnetic field in the $b^*\!c$ plane (Fig.~\ref{Fig2}(b)) as shown in Fig.~\ref{Fig2}(e). These results imply that the $g$-tensor is easy-axis ($g_c>g_{ab}$), assuming that at high temperature the susceptibility anisotropy primarily originates from the local physics of the $g$-tensor, as interactions have only subleading effects. The $g$-factor is independent of orientation in the $ab$ plane due to the $\bar{3}$ point group symmetry at the Ru sites.

Fig.~\ref{Fig3}(a) shows that there are major changes in the torque sign and shape as a function of temperature. The amplitude of the signal decreases upon cooling from 150~K (top curve), passes through zero, then changes sign at $T^*\simeq 95$~K and increases again in magnitude upon further cooling.
The torque sign change indicates that the stable equilibrium changes from $H\!\parallel\!c$ for $T>T^*$ to $H\!\parallel \!a$ for lower temperatures. A sign change in torque at the same temperature is also observed by measurements in the $b^*\!c$ plane
(see also Fig. S3 for sample S1 and Fig. S8 for sample S2 for two orientations in the SM
\cite{SM}). The observed sign change is attributed to the presence of anisotropic exchange interactions that favour a larger susceptibility for magnetic field applied in the $ab$ plane, parallel to the honeycomb layer ($\chi_{ab}>\chi_c$). The effects of the interactions is expected to become progressively more important upon cooling (Fig.~\ref{Fig3}(b)), and $T^*$ represents the temperature where anisotropy effects from the $g$-tensor and the interactions balance out (Fig.~\ref{Fig3}(g)). A sign change in torque as a function of temperature has also been observed in $\gamma$-Li$_2$IrO$_3$ and attributed to Kitaev interactions \cite{Modic2014}.

Below 30~K, the angular dependence of the torque for field in the $ac$ plane becomes progressively non-sinusoidal, as shown in Fig.~\ref{Fig3}(b).
The gradual development upon cooling of a non-sinusoidal, saw-tooth shape can be approximately parameterized by allowing a finite higher Fourier component $A_4\sin4(\theta-\theta_4)$ (Fig.~\ref{Fig3}(g)) or a direct fitting to an empirical expression  (described in Methods) using a saw-tooth amplitude, $A'$, and parameter $\gamma$. Thus, the development of a saw-tooth shape upon cooling, is illustrated by the rapid increase in the ratio $A_4/A_2$, plotted in Fig.~\ref{Fig3}(h) (filled circles), or the increase in the $\gamma$ parameter, shown in
shown Fig. S4(f) in the SM \cite{SM}.
We note that a similar saw-tooth torque shape, having the largest susceptibility for field in-plane, was also observed in $\alpha$-RuCl$_3$ \cite{Modic2018,Modic2021}. 
This has been attributed to an easy-plane magnetic anisotropy originating from an easy-plane $g$-tensor in combination with symmetric off-diagonal $\Gamma>0$ exchange \cite{Riedl2019}.
Despite of the observed changes in the shape of the torque signal,
the temperature dependence of
the different amplitudes do not show any anomalies and
have broadly Curie-Weiss dependencies
(see Fig. 3(g) and Fig. S4 in the SM \cite{SM}).
This confirms that RuI$_3$ does not show any signatures of long-range magnetic order, in agreement with susceptibility data \cite{Ni2022}.

\textbf{Magnetic anisotropy within the honeycomb $ab$ plane.} The magnetic anisotropy in the $ab$ plane (Fig.~\ref{Fig2}(c),(f),(i) and Fig.~\ref{Fig3}(c),(d) is much more challenging to probe as the torque signal is about 200 times smaller than that induced by the anisotropy between this plane and the $c$-axis. Thus, extracting the intrinsic in-plane anisotropy requires careful parameterization of other small contributions such as the finite sample weight and any small misalignment between the $c$-axis and the rotation axis
(see Fig.~S3 for sample S1 and Fig.~S8 for sample S2 in the SM \cite{SM}).
Fig.~\ref{Fig3}(c) shows the presence of a $\tau_6 = A_6\sin6(\phi-\phi_6)$ torque component, in addition to the $A_1$ and $A_{2,4}$ components attributed to the sample finite weight and out-of-plane axis misalignment, respectively. As expected for an admixed out-of-plane anisotropy component, $A_2$ changes sign at the crossing temperature $T^*$, as shown in Fig.~\ref{Fig3}(g). The resulting in-plane torque signal after subtracting the non-intrinsic $A_{1,2,4}$ contributions is shown in Fig.~\ref{Fig3}(d) and reveals a six-fold modulated signal with an amplitude that decreases rapidly upon heating and disappears above 20~K, as summarized in Fig.~\ref{Fig3}(h).

A six-fold modulated torque in the $ab$ plane is generically expected based on the crystal structure symmetry ($\bar{3}$ point group combined with time-reversal symmetry). The stable in-plane orientation deduced from Fig.~\ref{Fig3}(d) corresponds to a magnetic field close to the $b^*$-axis (the direction of Ru-Ru bonds), consistent with measurements on another sample S2
(see Fig.~S8(l) in the SM \cite{SM}).
 A {\em single} honeycomb layer has perpendicular mirror planes (the $ac$ plane and three-fold rotated versions, as shown in Fig.~\ref{Fig1}(a)), which constrain the free energy to have minima or maxima for an in-plane magnetic field either parallel or perpendicular to a Ru-Ru bond. The vertical stacking of layers in the $R\bar{3}$ structure
 (see Fig.~S2(a) in the SM \cite{SM})
  breaks those mirror planes, in principle lifting any restrictions on the in-plane equilibrium orientations. However, if inter-layer interactions are relatively much smaller than intra-layer, one would expect the stable equilibrium directions to still be close to those of isolated layers, consistent with the present measurements.

The in-plane torque signal cannot arise from single-ion effects, as Ru$^{3+}$ ($4d^5$) is a Kramers ion with an effective spin-$\frac{1}{2}$ ground state doublet that can have no local anisotropy, so the observed signal must be due to the cooperative effect of anisotropic spin interactions. A six-fold periodic torque was observed in $\alpha$-RuCl$_3$ \cite{Froude2024} below the ordering transition temperature and attributed to interactions that stabilize three-domain zigzag order. In the present case there is, however, no long-range magnetic order, suggesting a different mechanism for the observed effect.

\begin{figure}[htbp]
\centering
\includegraphics[trim={0cm 0cm 0cm 0cm},width=0.9\linewidth]{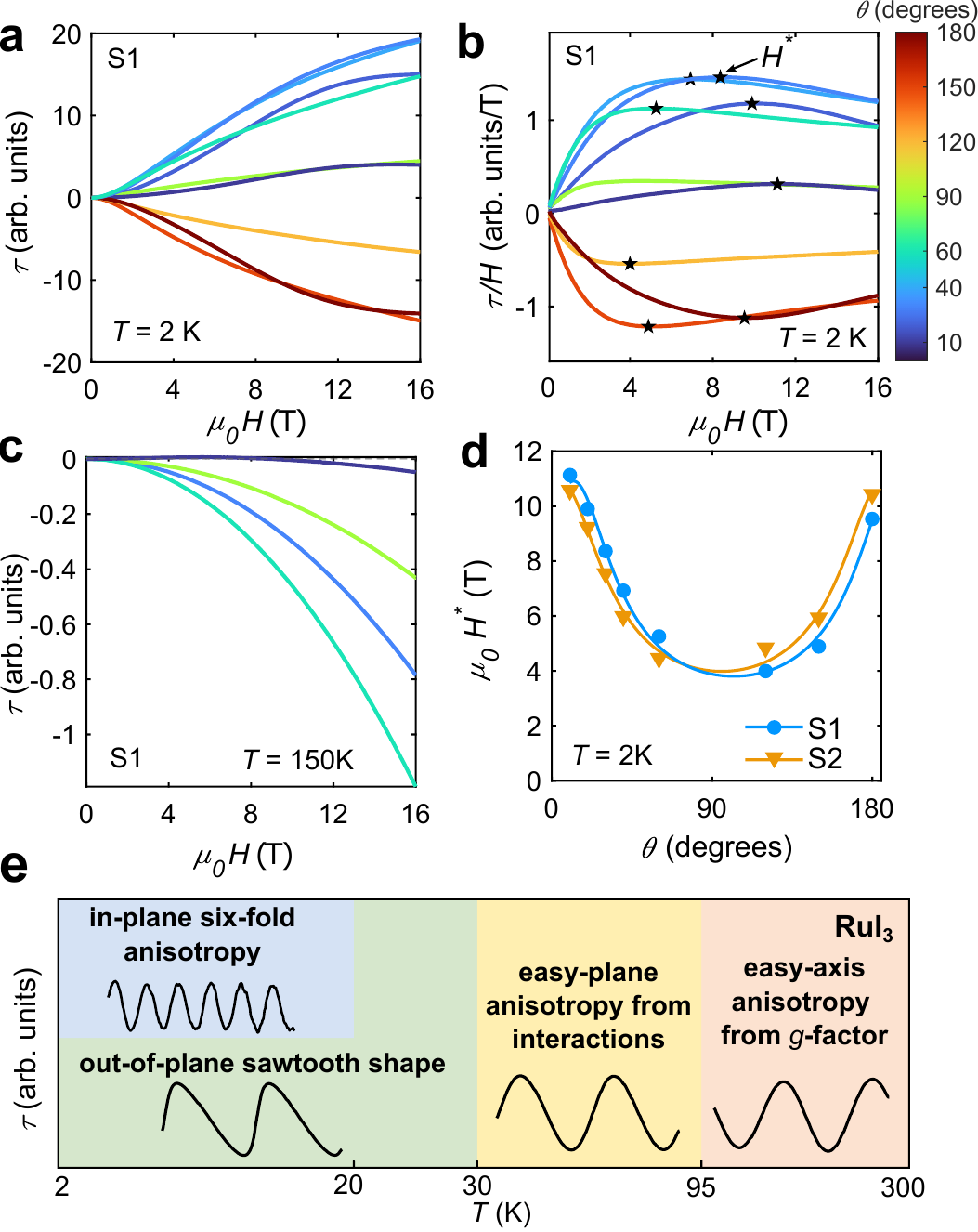}
\caption{
{\bf Field dependence of torque.}
(a) The field dependence of torque $\tau$ at 2~K is distinctly non-parabolic with a clear local maximum at a field $H^*$ in the scaled torque, $\tau/H$ versus $H$, in (b). In (a),(b) and (c) curves are colour-coded by the value of the field orientation angle $\theta$ in the $b^*\!c$ plane. (c) Field dependence of torque at 150~K shows a conventional parabolic $H^2$ dependence. (d) Angular dependence of the local maximum field $H^*$ for two different samples
(S1 and S2 from data shown in Fig. S10 in the SM cite{SM}).
Solid lines reflect an angular dependence parameterization, as described in the text. (e) Schematic representation of the changes in the torque as a function of temperature for RuI$_3$.}
\label{Fig4}
\end{figure}

{\bf Field-dependence of torque.} At high temperature (150~K) the torque has a parabolic dependence as a function of magnetic field strength, as expected for a paramagnet, and
shown in Fig.~\ref{Fig4}(c).  In contrast, at low temperatures  the torque is distinctly non-parabolic, as displayed in Fig.~\ref{Fig4}(a), with a clear broad maximum in the scaled torque $\tau/H$ versus $H$  at a field $H^*$ shown in Fig.~\ref{Fig4}(b). Previous field-dependent torque measurements in the Kitaev material $\gamma$-Li$_2$IrO$_3$ also observed a non-parabolic field dependence with a clear kink in $\tau/H$ at a field $H^*$ attributed to a phase transition between the low-field magnetically-ordered phase and a high-field quantum paramagnet \cite{Modic2014,Li2020}. In the present case there is no clear anomaly in the $\tau/H$ curve but only a broad maximum at $H^*$, as the low-field region does not have long-range magnetic order.
Thus, we attribute the field $H^*$ to a crossover in behaviour when the field strength matches a characteristic energy scale of interactions.
Empirically, the angular dependence of $H^*$ can be described by a phenomenological 
form for anisotropic systems,
as detailed in Methods.
This gives $H^*_c = 11(1)$~T and  $H^*_{ab} = 4.0(5)$~T, i.e. a field anisotropy factor $\alpha= H_c^*/H^*_{ab} = 2.7(5)$, in good agreement between different samples
in the planes of rotation normal to the
honeycomb layers
(see Table~S1 and Fig.~S10 in the SM \cite{SM}).
This field anisotropy factor is smaller than $\alpha \sim 10$ observed in $\gamma$-Li$_2$IrO$_3$ \cite{Modic2014}. Interestingly, the saw-tooth parameter, extracted
directly from the angular dependence of Fig.~\ref{Fig2}(g), shows a local maximum around 12~T, close to $H^*_c$, further supporting the scale of the field-induced effects
(see Fig.~S5(f)  for S1 and Fig.~S7(f) for S2 in the SM \cite{SM}).

{\bf Anisotropic exchange interactions.}
To understand the observed torque behaviour we consider an effective spin-$\frac12$ extended Kitaev Hamiltonian on the honeycomb lattice with the symmetry-allowed terms as given by the crystal structure of RuI$_3$. Here, previous \textit{ab-initio} studies \cite{Kaib2022} predicted dominant bond-dependent anisotropic interactions up to third nearest-neighbor and a strongly suppressed conventional Heisenberg exchange, suggesting a quantum spin liquid ground state. Torque calculations using the orthogonalized finite-temperature Lanczos method \cite{Morita2020} on the full \textit{ab-initio} model of Ref.~\cite{Kaib2022} (see SM \cite{SM}), capture the torque's six-fold anisotropy for in-plane orientations, and, for out-of-plane-orientation, the saw-tooth shaped torque at low temperatures and a sign-reversed sinusoidal torque at high temperatures. While these features agree well with experiment, the absolute sign of torque for this model is opposite to the one found experimentally. This overall sign difference can be accounted for in an \textit{adjusted} version of the \textit{ab-initio} parameters, in which in- and out-of-plane $g$-values are reversed, and the longer-range anisotropic coupling $\Gamma'_2$ is increased by 2\,meV (see SM \cite{SM}), compensating representatively for the likely underestimated longer-range couplings in the methodology of Ref.~\cite{Kaib2022}.
Based on this model, the observed saw-tooth torque shape in RuI$_3$ at low temperature described earlier can be attributed to strong in-plane anisotropy effects arising from the exchange interactions, which as confirmed by the observed sign change in torque, overcome the opposite-sign $g$-tensor anisotropy.
For this adjusted model, not only the measured relative angle- and temperature-dependence, but also the absolute sign of the torque is reproduced, as shown in Fig.~\ref{Fig3}(e) for field in the plane normal to the honeycomb layers and Fig.~\ref{Fig3}(f) for field in the honeycomb layers. Furthermore, the susceptibility for $H\!\parallel\! b^*$ is larger than that for $H\!\parallel\! c$ at low temperatures, with a reversed order at high temperatures, as shown in Fig.~\ref{Fig3}(i) and deduced experimentally from Fig.~\ref{Fig3}(g). The experimentally-observed angle- and temperature-dependent torque behaviour is therefore attributed to the presence of strongly anisotropic and long-ranged (up to third nearest neighbor) magnetic exchange interactions.

It is interesting to compare the observed torque behaviour in RuI$_3$ -- schematically summarized in Fig.~\ref{Fig4}(e) -- to that experimentally found in isostructral $\alpha$-RuCl$_3$ \cite{Leahy2017,Modic2018,Froude2024} and theoretically discussed in \cite{Riedl2019}, which in contrast to RuI$_3$ shows well-established zigzag order at low temperatures. In both materials, at low temperatures the torque for field rotated in the plane normal to the honeycomb layers has two-fold periodicity and saw-tooth shape (for $\alpha$-RuCl$_3$ only at rather high fields) with the stable field orientation in-plane \cite{Leahy2017,Riedl2019}, suggesting that bond-dependent interactions in both cases create easy-plane type anisotropy. The torque for in-plane rotated field displays six-fold periodicity with the stable field orientation near the Ru-Ru bond direction in both cases, with a sinusoidal shape in RuI$_3$, but pronounced saw-tooth shape in $\alpha$-RuCl$_3$ \cite{Froude2024}
and $\alpha$-RuBr$_3$ \cite{Pearce2024}, which may be related to the fact that RuI$_3$ has only short-range correlations whereas $\alpha$-RuCl$_3$ has static long-range order. In $\alpha$-RuCl$_3$ clear anomalies occur in the torque as a function of increasing field when the zigzag order is suppressed (by in-plane magnetic fields of 7-8~T) \cite{Leahy2017}. RuI$_3$ shows no clear anomalies as a function of increasing field strength for any orientation, but a broad maximum at an angle-dependent field $H^*(\theta)$ in the scaled torque $\tau(\theta)/H$ as a function of field $H$, which we attribute not to a phase transition, but to a crossover in behaviour when the field strength matches a characteristic energy scale of interactions.

In contrast to $\alpha$-RuCl$_3$, which is located deep inside the Mott insulating regime,  ARPES measurements of RuI$_3$ show evidence of bands crossing the Fermi level \cite{Louat2023}, whereas  transport and NMR data resemble those of a bad metal \cite{Ni2022, Nawa2021}. RIXS data show evidence for proximity to a bandwidth-controlled metal-to-insulator transition \cite{Gretarsson2024},
also supported by \textit{ab initio} calculations \cite{Kaib2022}.
The present observation of a strong angular dependence of the magnetic torque,
in turn, implies a highly-structured angular dependence of the magnetic free energy, which is difficult to explain in a conventional metallic picture.
Instead, this behaviour arises naturally as a result of strongly anisotropic bond-dependent magnetic interactions between spin-orbit entangled $j_{\rm eff}=1/2$ localized Ru$^{3+}$ magnetic moments described by extended Kitaev models.
Therefore, RuI$_3$ emerges as a novel candidate Kitaev material where the close proximity to an insulator to metal transition generates substantial longer-range frustrated magnetic interactions and enhanced fluctuations that could stabilize unconventional cooperative magnetism. Our findings open a new direction of probing and understanding magnetism in correlated systems on the verge of an insulator-to-metal transition.

\newpage

\vspace{0.2cm}

{\bf Methods}

\textbf{Structural characterization of samples.} Torque measurements were performed on several hexagonal-shaped single crystals of RuI$_3$ of typical diameter 100~$\mu$m and thickness 50~$\mu$m, isolated from a polycrystalline batch, synthesized as reported in \cite{Ni2022}. Single crystal x-ray diffraction measurements performed using a Mo source SuperNova diffractomer confirmed all selected samples were un-twinned single crystals, with sharp Bragg peaks and no detectable diffuse scattering indicating absence of stacking faults, and with an x-ray diffraction pattern that agreed with the nominal rhombohedral $R\bar{3}$ crystal structure with a 3-layer $ABC$ stacking sequence proposed in \cite{Ni2022}
(see Fig.~S2 in the SM \cite{SM}).
 We note that a related polymorph of RuI$_3$ with a 2-layer stacking sequence ($P\bar{3}1c$ space group), was also reported using a different synthesis protocol \cite{Nawa2021,Gretarsson2024}, those two structures are easily distinguishable by x-ray studies and all samples in the present study had the 3-layer $R\bar{3}$ structure. We label crystal axes in the hexagonal setting and choose as reference reciprocally-orthogonal directions the in-plane hexagonal $a$-axis, the in-plane $b^*$-axis orthogonal to $a$, and $c$-axis normal to the honeycomb layers, as illustrated in Fig.~\ref{Fig1}(a).

\textbf{Torque experiments.} For the torque measurements each sample was mounted on a Seiko PRC400 piezocantilever using vacuum grease. In an applied magnetic field $\bm{H}$ the sample is subject to a torque $\bm{\tau}=\mu_0\bm{m}\times\bm{H}$,  where $\bm{m}$ is the magnetic moment of the sample. The torque projection $\tau$ onto the cantilever rotation axis tends to bend the cantilever creating mechanical stress at its base, detected via a voltage change
in a Wheatstone bridge circuit.
There is a linear relation between the torque and the voltage change in the limit of small cantilever deviations, which is the case for all measurements reported here, and the torque detection sensitivity is of the order of $10^{-13}$~Nm. To obtain the angular dependence of the torque, the cantilever platform was rotated a full 360$^{\circ}$ around a rotation axis normal to the (fixed) applied field direction as illustrated in Fig.~\ref{Fig2} (top row). The sample was mounted in three different orientations in order to probe the torque for the magnetic field relative to the crystallographic axes in three orthogonal planes $ac$, $b^*\!c$ (using $\theta$ angle) and $ab$ (using $\phi$ angle). Results are reproducible between all measured samples and most data presented here are for sample S1
(the results for sample S2 are shown in Figs. S2, S6, S8, S7, S9, and S10
in the SM \cite{SM}).
To ensure measurements are quantitatively comparable between the three different sample orientations, the sample was mounted in each case in approximately the same centre-of-mass position relative to the cantilever. Measurements were performed using a Quantum Design PPMS system in the temperature range from 300~K to 2~K and magnetic fields up to 16~T. Data were collected as angular dependencies both for clockwise and anticlockwise directions at fixed field (all plotted data are for the anticlockwise direction) and temperature, as well as at base temperature for a set of selected fixed orientations as a function of applied field strength. Angular dependencies of torque in zero field were used to estimate the non-magnetic torque effect due to the finite sample weight (see the lowest curves in Fig.~\ref{Fig2}(f) and (i)) and to set the absolute sign of the torque using the convention described below.

\textbf{Torque sign convention.} Throughout we plot the torque projection $\tau$ onto the cantilever rotation axis as the sample platform is rotated by a variable angle $\theta$ around this axis oriented normal to the (fixed) applied magnetic field direction, with positive sign for $\tau$ in the sense of increasing the rotation angle $\theta$. $\tau$ is then equal to the negative of the angular derivative of the free energy $\tau=-\frac{\partial F}{\partial \theta}$ with stable equilibrium (local minimum in the free energy $F$) at an angle
${\theta_0}$
corresponding to torque passing through zero with a negative gradient,
i.e. $\tau({\theta_0})=0$ and $\left.\frac{\partial\tau}{\partial\theta}\right|_{\theta={\theta_0}}<0$, a restoring torque of the form $\tau(\theta)=-k(\theta-{\theta_0})$ with $k>0$ for small deviations.
Consequently, unstable equilibrium corresponds to torque passing through zero with a positive gradient.  Note that this sign convention (\textit{negative} angular derivative of the free energy) differs from the one chosen in, e.g., Refs.~\cite{Riedl2019,Froude2024}.

{\bf Torque angular dependence parameterisation.} We parameterise the torque angular dependence using the Fourier decomposition
$$\tau(\theta) = \tau_1(\theta-\theta_1)+\sum_{n=2,4,6}\tau_n(\theta-\theta_n),$$
where the first term, $\tau_1(\theta)=A_1 \cos \theta$, accounts for the effect of the sample weight, with 360$^\circ$ periodicity and $\theta_1$ is a small experimental angular offset.
The rest of the terms, $\tau_n(\theta)=A_{n} \sin n \theta$,
with $n=2,4,6$ arise from magnetic anisotropy effects,
$A_n$ are the different Fourier amplitudes and $\theta_{n}$ are small angular offsets.
In zero field only the weight effect, $\tau_1$, contributes to the total torque. For magnetic field in the plane normal to the honeycomb layers we assume $\tau_6=0$. For magnetic field in the $ab$ plane the relevant rotation angle is $\phi$, instead of $\theta$, and all above Fourier components are included, where $\tau_{2}$ and $\tau_{4}$ are attributed to the admixture of out-of-plane anisotropy due to a small misalignment between the $c$-axis and the normal to the rotation plane, and $\tau_6$ is the intrinsic in-plane six-fold component.

{\bf The sawtooth parameterisation of torque data.}
To characterize the non-sinusoidal, saw-tooth shape of the torque angular dependence observed at low temperatures for field in the plane normal to the honeycomb layers we also use the alternative parameterization by an empirical saw-tooth form $\tau(\theta)=\tau_1(\theta-\theta_1)+\tau'(\theta-\theta')$ with the intrinsic term
$$\tau'(\theta)= A'\frac{1+\gamma}{2} \frac{\sin 2 \theta }{\sqrt{\cos^2 \theta + \gamma^2 \sin^2 \theta}},$$
where $A'$ is the intrinsic torque amplitude and the saw-tooth parameter $\gamma$ is the ratio of the absolute gradients when $\tau'$ crosses zero with a positive/negative gradient and $\theta'$ is a small experimental angular offset. For an intrinsic sinusoidal torque shape $\gamma=1$,  $\tau'(\theta)=\tau_2(\theta)$, $A'=A_2$ and $\tau_4=0$.
An example
of the fitting procedure is shown in Fig.~S11 and the
different extracted parameters are shown in the Figs.~S4, S5,
S7 and S9, in the SM \cite{SM}.

{\bf The angular dependence parameterisation of $H^*$.}
We parameterise the angular dependence of
$H^*$ using
 a phenomenological form for anisotropic systems,
 $$\left(\frac{H^* \cos (\theta-\theta^*)}{H^*_{c}}\right)^2+ \left(\frac{H^* \sin (\theta-\theta^*)}{H^*_{ab}}\right)^2=1.$$
 This enables to
 extract two characteristic fields,
 $H^*_c$ and  $H^*_{ab}$, and
 a field anisotropy factor $\alpha= H_c^*/H^*_{ab}$
($\theta^*$ is an angular offset), as detailed in
Table~S1 in the SM \cite{SM}.

{\bf Theoretical calculations.}
To calculate torque and magnetic susceptibility for the considered extended Kitaev model, we use both standard exact diagonalization for ground state ($T=0$) properties, and the orthogonalized finite-temperature Lanczos method (OFTLM) introduced in Ref.~\cite{Morita2020}
for finite-temperature ($T>0$), both on a 24-site periodic cluster that represents all point group symmetries of the lattice. Following the notation of Ref.~\cite{Morita2020},
we employ for OFTLM $N_v=1$ exact eigenstate, $M=80$-dimensional Krylov-subspaces, and at least $R\ge 20$ sampling vectors per considered Hamiltonian until the desired level of convergence is reached.
To analyze the latter, statistical errors were estimated using standard jackknife resampling methods. For all plotted OFTLM results, the estimated standard deviation of the results is smaller than the graphical width of the plotted line
(see Fig. S1 in the SM \cite{SM}).

{\bf Competing interests.} The authors declare no competing interests.

{\bf Additional information.} Supplementary Materials are available for this paper at \cite{SM}.

{\bf Acknowledgements.} ZM is grateful to Hertford College Oxford for funding. We thank Jay Patel and Ioana Paulescu for their earlier contribution to the preliminary studies. This work was partially supported by the EPSRC (EP/I004475/1), the Oxford Centre for Applied Superconductivity
and the ISABEL project of the European Union’s Horizon 2020 research
and innovation programme under grant agreement No 871106. We acknowledge financial support from the Oxford University John Fell Fund. AIC acknowledges an EPSRC Career Acceleration Fellowship (EP/I004475/1). KM and RC acknowledge support from the European Research Council under the European Union’s Horizon's 2020 Research and Innovation Programme Grant Agreement Number 788814 (EQFT). DK and RV acknowledge support by the Deutsche Forschungsgemeinschaft (DFG, German Research Foundation) for funding through TRR 288-422213477 (project A05, B05).  RV, RC and AIC are grateful for hospitality by the Kavli Institute for Theoretical Physics (KITP) where part of this work was discussed, supported in part by the National Science Foundation under Grants No. NSF PHY-1748958 and PHY-2309135. The work at Princeton University was supported by the Gordon and Betty Moore Foundation grant number GMBF-9066.

\bibliography{RuI3_bib}

\end{document}